\documentclass[10pt,preprint]{aastex}

\newcommand{\der}[2][\;\;]{\ensuremath{ \frac{d{#1}}{d{#2}} }}
\newcommand{\dern}[3][\;\;]{\ensuremath{ \frac{d^{#3}{#1}}{d{#2}^{#3}} }}
\newcommand{\dpar}[2][\;\;]{\ensuremath{ \frac{\partial{#1}}{\partial{#2}} }}
\newcommand{\dparn}[3][\;\;]{\ensuremath{ \frac{\partial^{#3}{#1}}{\partial{#2}^{#3}} }}

\newcommand{\e}{{\rm e}}

\newcommand{\bvec}[1]{{\mbox{{\boldmath$#1$}}}} 

\newcommand{\eqnref}[1]{(\ref{#1})}

\def\figone{%
\begin{figure*}[t]%
        \epsscale{0.4}%
        \plotone{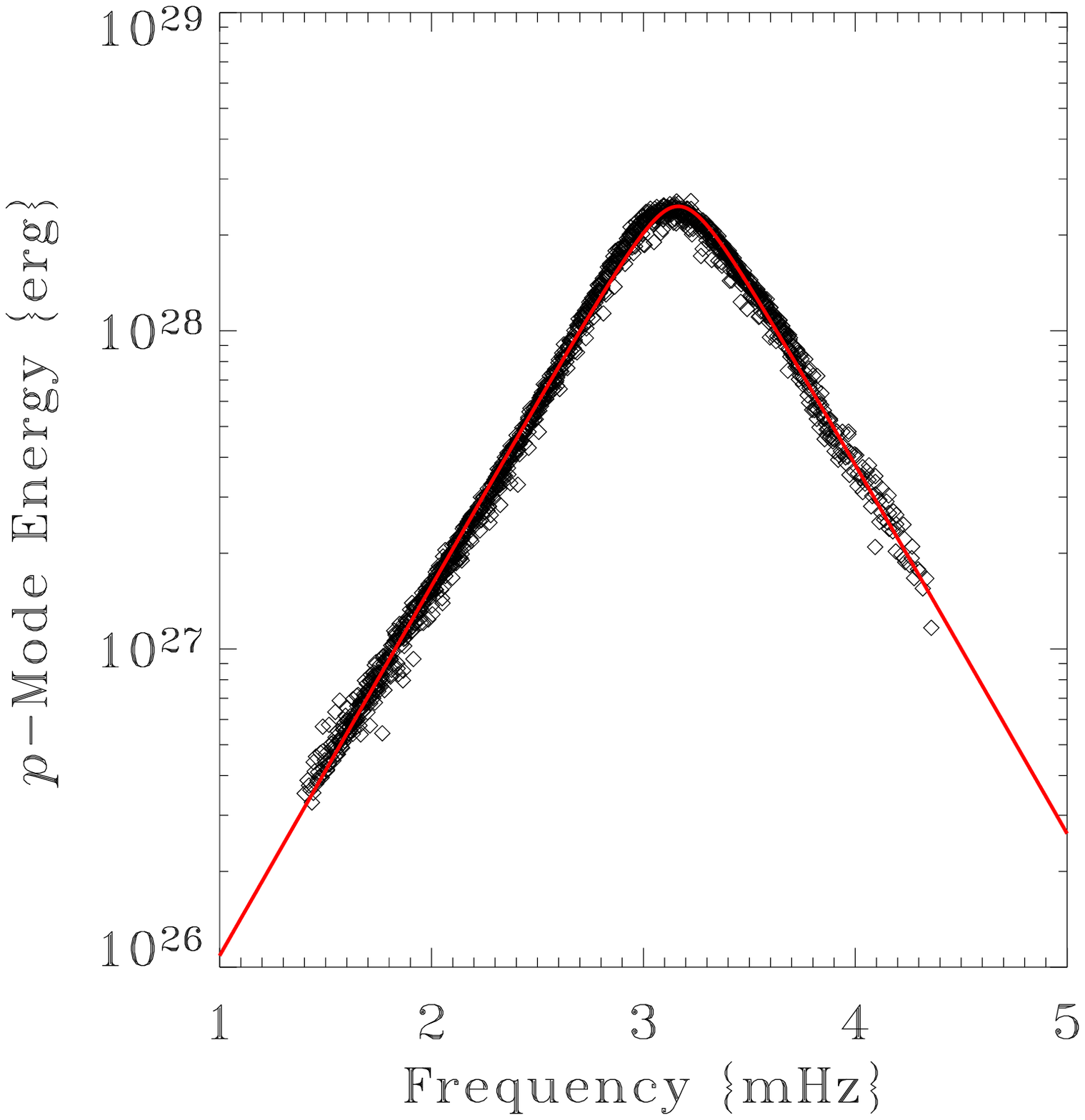}%
        \caption{\small Energy contained in solar $p$ modes with harmonic
degrees in the range $3<\ell\leq 150$ and orders $n>0$. Diamonds indicate the
measurements of \cite{Komm:2003}. As can be clearly seen, the energy is primarily
a function of frequency. The solid red curve is a maximum-likelihood fit to the
data using the functional form appearing in eq.~\eqnref{eqn:Efit}.%
\label{fig:modeenergy}}%

\end{figure*}%
}

\def\figtwo{%
\begin{figure*}[t]%
        \epsscale{0.8}%
        \plotone{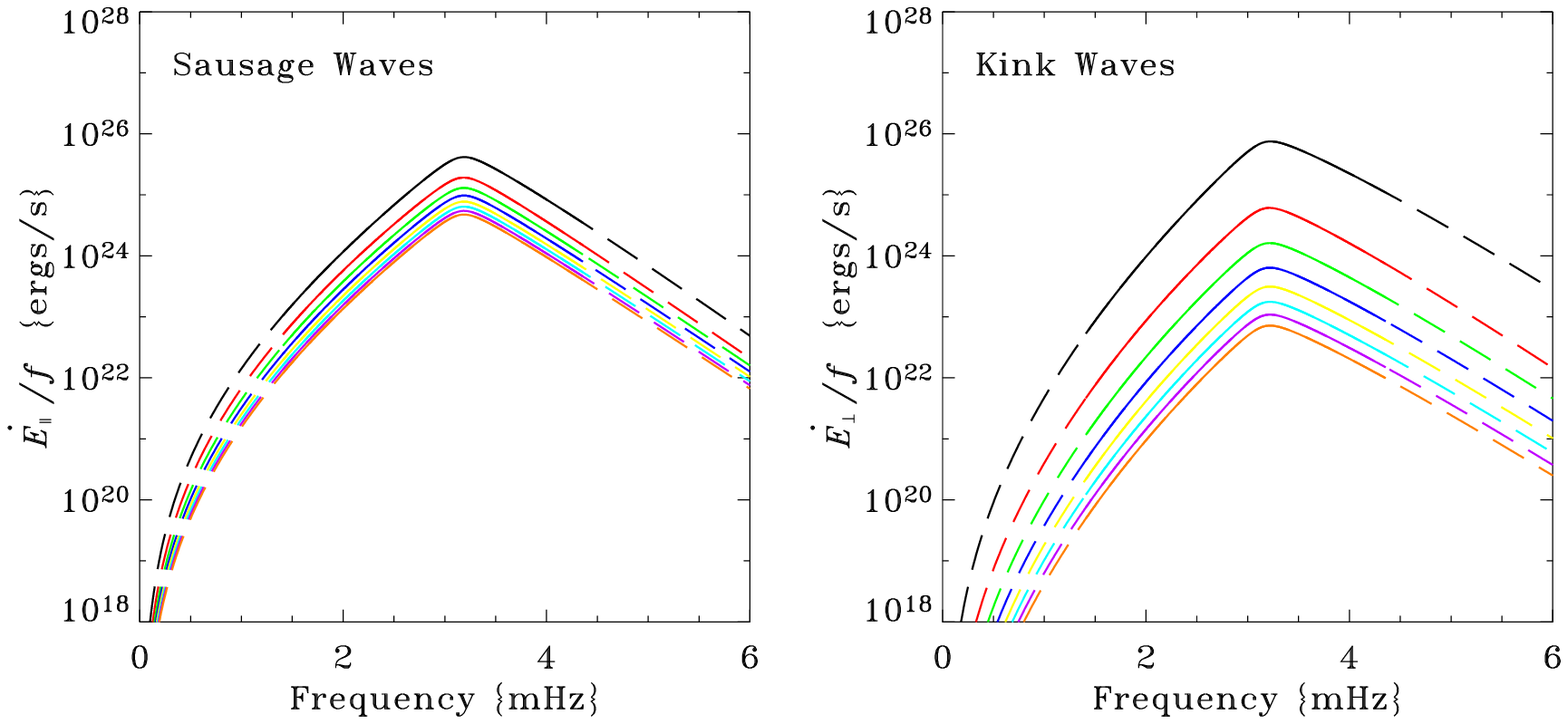}%
        \caption{\small Energy flux of tube waves driven up through the model
photosphere.  The solar surface is peppered with identical thin flux tubes with
$\beta = 1.0$ and a combined filling factor $f$. The fluxes shown are divided by
this filling factor. The different curves correspond to the flux driven by different
order $p$ modes: $f$ ({\it black}), $p_1$ ({\it red}), $p_2$ ({\it green}), $p_3$ ({\it dark blue}),
$p_4$ ({\it yellow}), $p_5$ ({\it aqua}), $p_6$ ({\it violet}), and $p_7$ ({\it orange}). The range of frequencies
for which the curves are solid corresponds to the window where \cite{Komm:2003}
measured $p$-mode energies. Frequencies where the curves are dashed indicate
extrapolations of the measured $p$-mode energy. These curves are for the case where
a maximal-flux boundary condition has been applied at the photosphere.%
\label{fig:Eatmo}}%

\end{figure*}%
}

\def\figthree{%
\begin{figure*}[t]%
        \epsscale{0.4}%
        \plotone{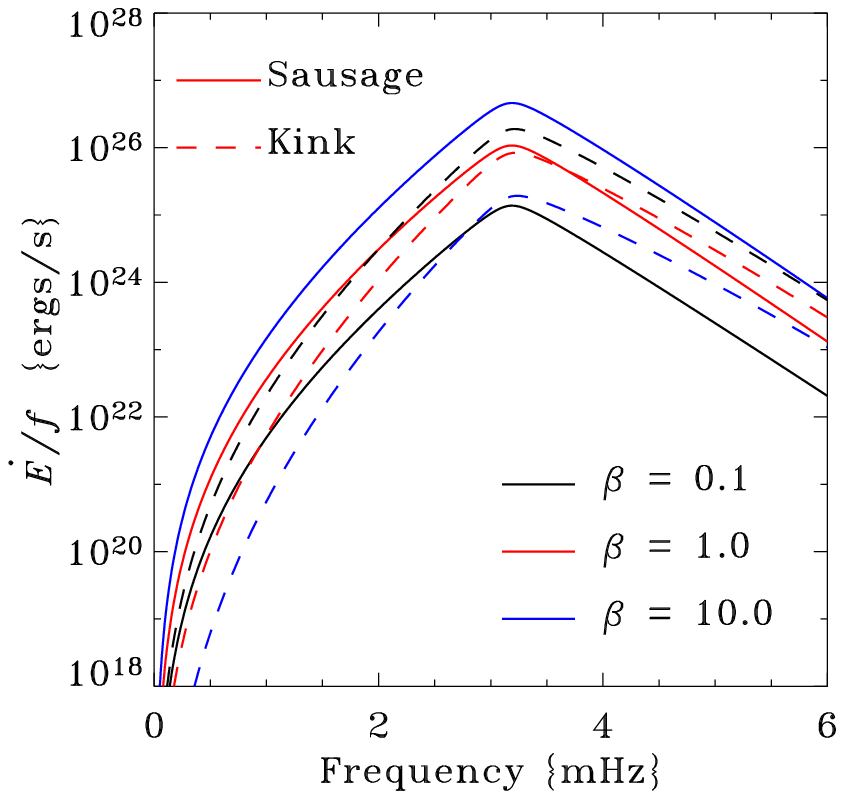}%
        \caption{\small Energy flux of tube waves driven up through the model
photosphere summed over all mode orders $n<8$. The different colors correspond to
models with flux tubes with different values of $\beta$: $\beta=0.1$ ({\it black}),
$\beta = 1.0$ ({\it red}), and $\beta = 10.0$ ({\it blue}).
\label{fig:Eatmotot}}%

\end{figure*}%
}

\def\figfour{%
\begin{figure*}[t]%
        \epsscale{0.8}%
        \plotone{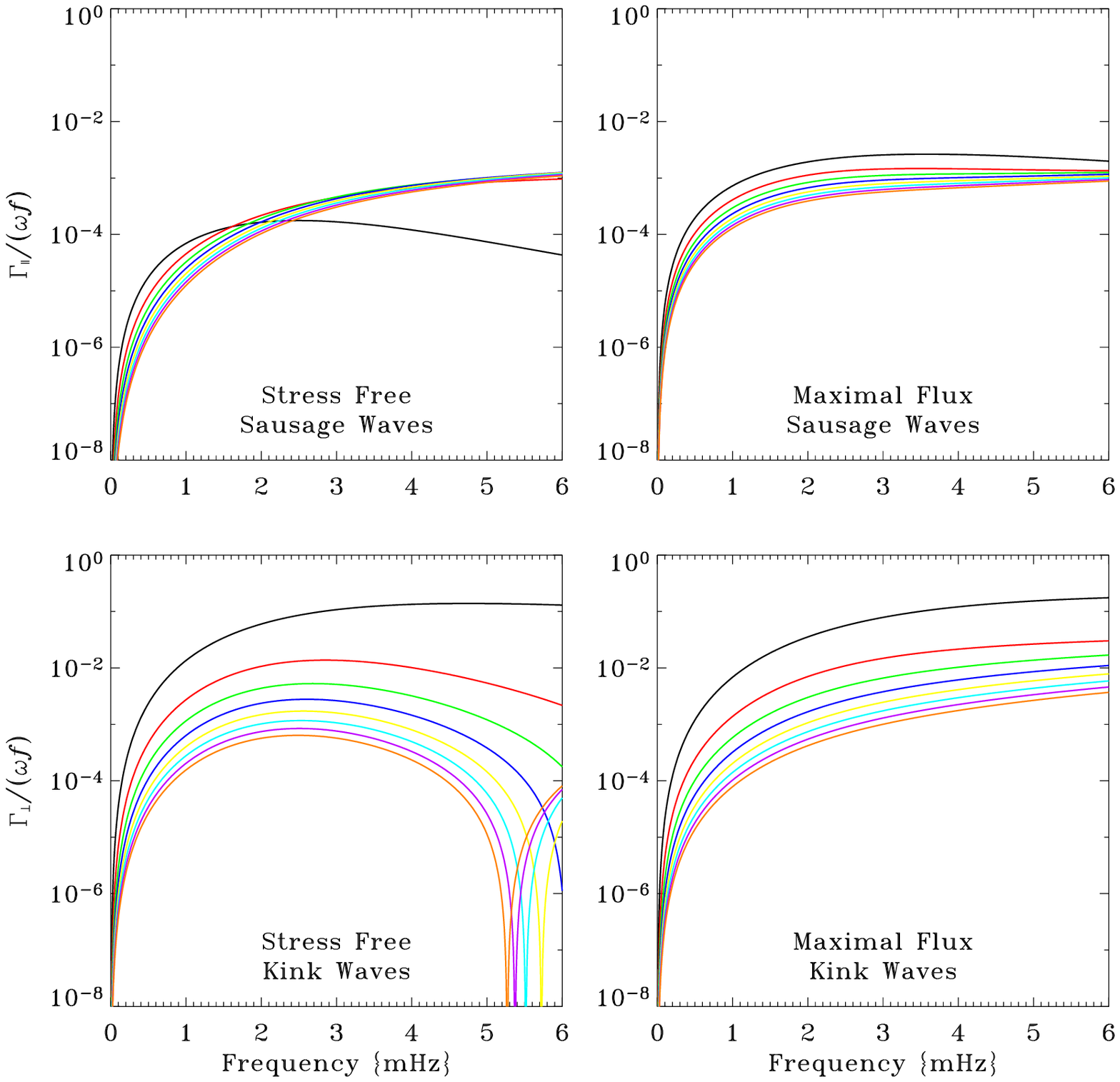}%
        \caption{\small Damping rate of $f$ and $p$ modes caused by the excitation
of both sausage and kink waves on thin magnetic flux tubes. The tubes have $\beta = 0.1$.
Two different photospheric boundary conditions are applied: stress free and maximal flux.
The nulls in the damping rate seen between 5 and 6 mHz for the kink wave under the stress-free
boundary condition arise because the downward-propagating wave and the wave that reflects
off the upper surface destructively interfere.
\label{fig:gamma_beta0.1}}%

\end{figure*}%
}

\def\figfive{%
\begin{figure*}[t]%
        \epsscale{0.8}%
        \plotone{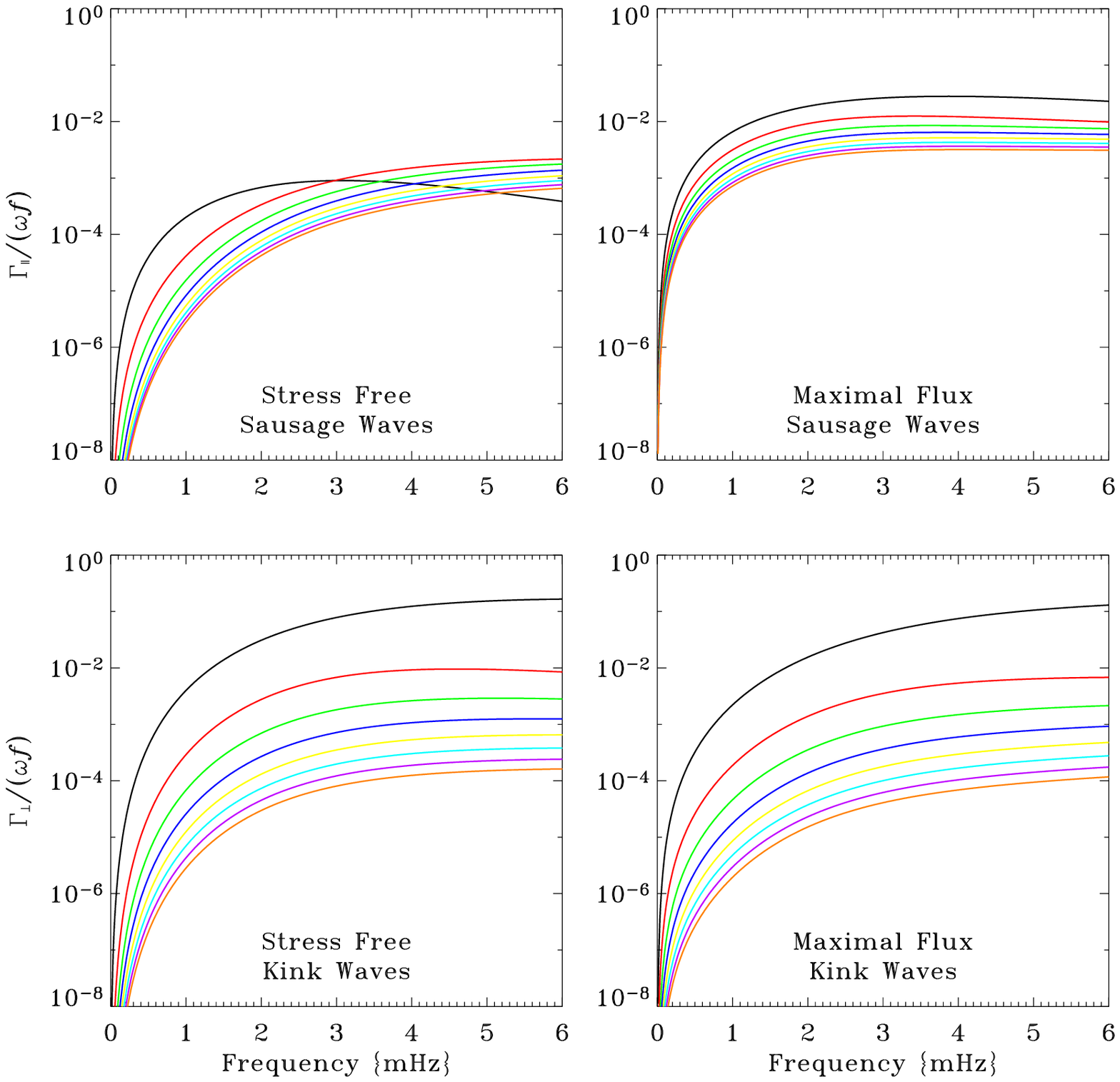}%
        \caption{\small Same as Fig.~\ref{fig:gamma_beta0.1}, except that the thin flux tubes have $\beta=1.0$.%
\label{fig:gamma_beta1.0}}%

\end{figure*}%
}

\def\figsix{%
\begin{figure*}[t]%
        \epsscale{0.8}%
        \plotone{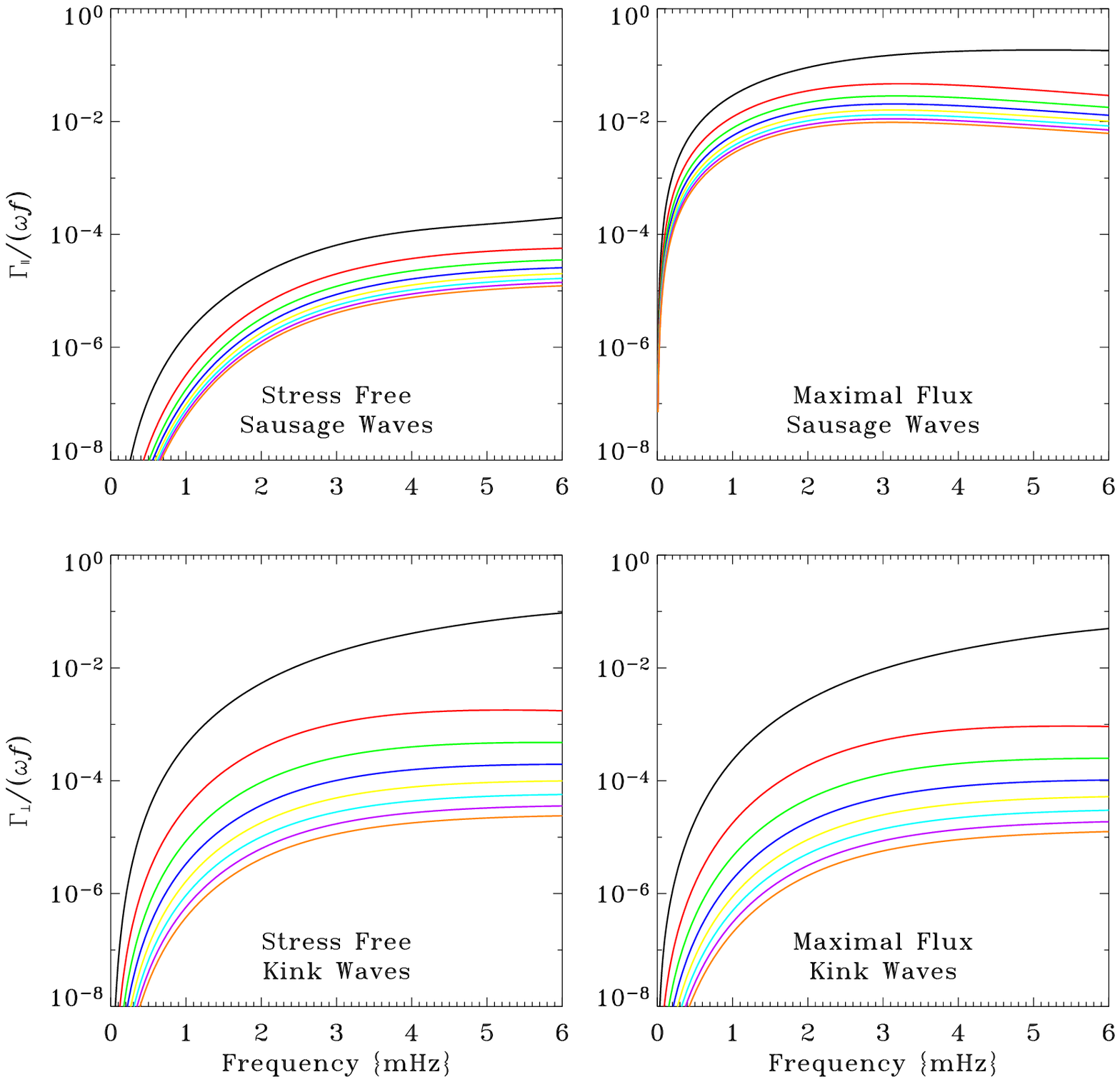}%
        \caption{\small Same as Fig.~\ref{fig:gamma_beta0.1}, except that the thin flux tubes have $\beta=10.0$.%
\label{fig:gamma_beta10}}%

\end{figure*}%
}

\def\figseven{%
\begin{figure*}[t]%
        \epsscale{0.8}%
        \plotone{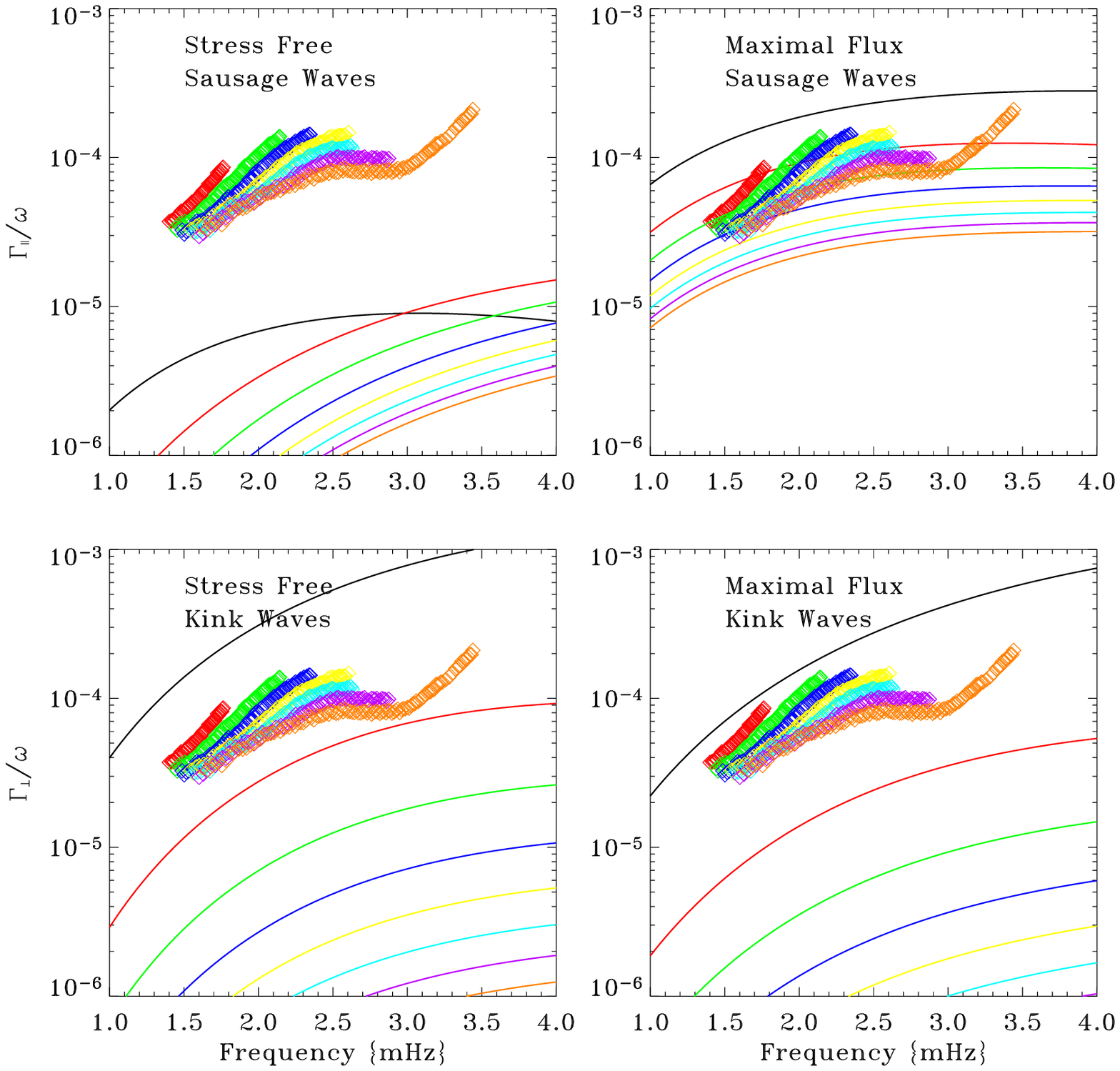}%
        \caption{\small Same as Fig.~\ref{fig:gamma_beta0.1}, except that the
value of $\beta$ is $1.0$ and measured full widths of the $p$ modes have been
overplotted with diamonds. Each mode order is plotted with a different color, using
the same color scheme as in the theoretical curves: $f$ ({\it black}), $p_1$ ({\it red}), etc. The
full widths were measured by \cite{Komm:2003} and only orders 1--7 are shown.%
\label{fig:gamma_data}}%

\end{figure*}%
}

\begin{document}

\title{The Generation of Coronal Loop Waves below the Photosphere by $\bvec{p}$-Mode Forcing
\footnote{This paper was originally published in the Astrophysical Journal on 10 April 2008, vol. 677, pp. 769--780}}

\author{Bradley W. Hindman}
\affil{JILA and Department of Astrophysical and Planetary Sciences,
University of Colorado, Boulder, CO~80309-0440, USA}
\email{hindman@solarz.colorado.edu}
\author{Rekha Jain}
\affil{Applied Mathematics Department, University of Sheffield, Sheffield S3 7RH, UK}


\begin{abstract} 

Recent observations of coronal-loop waves by {\it TRACE} and within the corona as a
whole by CoMP clearly indicate that the dominant oscillation period is 5 minutes, thus
implicating the solar $p$ modes as a possible source. We investigate the generation
of tube waves within the solar convection zone by the buffeting of $p$ modes. The
tube waves---in the form of longitudinal sausage waves and transverse kink waves---are
generated on the many magnetic fibrils that lace the convection zone and pierce the solar
photosphere.  Once generated by $p$-mode forcing, the tube waves freely propagate
up and down the tubes, since the tubes act like light fibers and form a waveguide
for these magnetosonic waves. Those waves that propagate upward pass through the
photosphere and enter the upper atmosphere where they can be measured as loop oscillations
and other forms of propagating coronal waves. We treat the magnetic fibrils as
vertically aligned, thin flux tubes and compute the energy flux of tube waves that can generated
and driven into the upper atmosphere. We find that a flux in excess of $10^5$
ergs cm$^{-2}$ s$^{-1}$ can be produced, easily supplying enough wave energy to explain the
observations. Furthermore, we compute the associated damping rate of the driving
$p$ modes and find that the damping is significant compared to observed line widths
only for the lowest order $p$ modes.

\end{abstract}



\section{Introduction}
\label{sec:introduction}
\setcounter{equation}{0}

It is now well established through observations by the {\sl Solar and Heliospheric
Observatory} ({\it SOHO}), the {\it Transition Region and Coronal Explorer} ({\it TRACE}), and {\it Hinode}
that the solar atmosphere supports the propagation of a variety of magnetohydrodynamic
(MHD) waves. Slow magnetosonic waves have been implicated in the observations of
coronal plumes \citep{Ofman:1997, Ofman:1999, DeForest:1998} and in coronal-loop
oscillations \citep[e.g.,][]{Berghmans:1999, DeMoortel:2000, DeMoortel:2002a}.
\cite{Wang:2002} interpreted oscillatory behavior seen in very hot ($>$ 5 MK) coronal
loops by {\it SOHO}/SUMER, as standing slow magnetosonic waves of long wavelength.  A
propagating intensity oscillation with 5 minute periodicity in a ``nonsunspot"
coronal loop has also been reported by \cite{Marsh:2003}. They estimate a speed of
50--195 km s$^{-1}$ for this wave and suggest, on the basis of Coronal Diagnostic
Spectrometer ({\sl SOHO}/CDS) data, that it is also present at chromospheric, transition
region, and coronal temperatures.

Fast magnetosonic waves have also been invoked to explain coronal and chromospheric
oscillation observations.  \cite{Williams:2002} report intensity fluctuations with
a period of 6 s along and within coronal loops. The propagation speed is estimated
to be 2 $\times$ 10$^{3}$ km s$^{-1}$, suggesting the possibility of fast magnetosonic waves.
\cite{Verwichte:2005} have recently claimed to have seen propagating fast kink waves
in a postflare supra-arcade. Transverse loop oscillations observed by {\it TRACE}
\citep{Aschwanden:1999, Nakariakov:1999} have also been linked to fast kink waves.
Finally, there have been suggestions that the large-scale EIT waves discovered by
\cite{Thompson:1998} are the result of propagating fast magnetosonic waves
\citep{Wang:2000, Murawski:2001, Ofman:2002}.

Recent observations by the Coronal Multichannel Polarimeter (CoMP) instrument
\citep{Tomczyk:2007} and {\it TRACE} \citep[see the overview by][]{DeMoortel:2007}
clearly indicate that coronal oscillations are ubiquitous and have amplitudes
that peak at the $p$-mode frequency band.  In the high corona, the CoMP instrument
detects strong fluctuations of the line-of-sight Doppler velocity with a broad
peak at 3.5 mHz. Correspondingly weak intensity oscillations are seen at the same
frequency. This indicates that the observed waves are nearly incompressive, and
as such \cite{Tomczyk:2007} identified the waves as Alfv\'enic. However, in a
low-$\beta$ plasma, fast magnetosonic tube waves or kink waves are also nearly incompressive.
Therefore, we posit that these observations in the high corona may also be fast
kink waves.

On the other hand, the loop oscillations observed by {\it TRACE} have been identified
as slow waves \citep[e.g.,][]{Berghmans:1999,DeMoortel:2000, DeMoortel:2002a}.
The oscillations associated with sunspot loops have peak power at a period of
3 minutes (or a frequency of 5.5 mHz), while loops not associated with sunspots
have peak power at a period of 5 minutes \citep[or 3.3 mHz; ][]{DeMoortel:2002b, Centeno:2006}.
Therefore, both slow and fast waves with frequencies associated with $p$-mode
oscillations have been observed. The $p$ modes, and possibly solar granulation,
are clearly implicated as the ultimate source of many coronal fluctuations.
In this paper, we concentrate on $p$ modes as the driving mechanism.

If $p$ modes are in fact responsible, the $p$ modes must somehow manage to leak
into the upper atmosphere, despite the fact that they become evanescent above the
photosphere.  Some have suggested that the leakage is accomplished by a ``ramp effect,"
by which an inclined magnetic field effectively reduces the acoustic cutoff frequency
of the atmosphere \citep{DePontieu:2005, Hansteen:2006, McIntosh:2006, Jefferies:2006}.
However, we point out that the cutoff frequency for both the slow sausage wave
and the fast kink wave can fall below the nonmagnetic acoustic cutoff frequency 
(see \citeauthor{Roberts:1978} [\citeyear{Roberts:1978}] and \citeauthor{Musielak:2001} [\citeyear{Musielak:2001}]
for a derivation of the cutoff for sausage and kink waves, respectively).
Therefore, a flux tube can act as a waveguide \citep{Roberts:1981},
permitting efficient penetration of the acoustic barrier presented by the photosphere.
It may be possible that vertically aligned flux tubes are capable of carrying a
sufficient flux of tube waves into the atmosphere to explain the observations.

We suggest the following mechanism. Within the solar convection zone, $p$-mode
oscillations buffet the many magnetic fibrils that thread through the photosphere
into the chromosphere and corona. This buffeting excites both sausage and kink waves
on the fibrils, which then propagate both up and down the vertical field lines. Those
waves that travel downward are lost in the convection zone, whereas those that
propagate upward pass through the photosphere into the upper atmosphere, where they
manifest as coronal-loop oscillations and upward-propagating waves.

In this paper we estimate, by a semianalytic method, the atmospheric energy flux
of sausage and kink waves that can be generated by this mechanism. Our calculation
is a direct extension of the work by \cite{Bogdan:1996}, hereafter referred to as BHCC96.
We demonstrate that $f$ and $p$ modes efficiently generate tube waves and that an energy
flux in excess of $10^{5}$ ergs cm$^{-2}$ s$^{-1}$ can be driven upward through photospheric levels.
Whether, such a flux can survive passage through the chromosphere and transition region
into the corona is left to subsequent work.

The paper is organized as follows. In \S 2 we describe the stationary configuration of
the flux tubes and the equilibrium atmosphere in which they are embedded.  \S 3 details
the $p$-mode oscillations, and \S 4 derives the driven oscillations of the thin flux tubes.
In \S 5 we present our primary results, the energy flux of tube waves and the damping
rates of the $p$ modes that those fluxes engender.  \S 6 provides a discussion of our
findings, where we compare our derived atmospheric energy fluxes with those obtained by
observations.


\section{The Equilibrium}
\label{sec:equilibrium}
\setcounter{equation}{0}

We examine the excitation of waves on a thin magnetic fibril through the buffeting
of the tube by acoustic oscillations within the solar convection zone. The fibril
threads through a field-free atmosphere, and its axis is aligned with the force of
gravity. We assume that in the absence of the acoustic wave field, the fibril and
surrounding atmosphere are static and in equilibrium. In the following two subsections
we describe the field-free atmosphere in which the fibril is embedded and the
equilibrium configuration and structure of the fibril itself.

\subsection{The Atmosphere}
\label{subsec:equil_atmo}

We restrict our attention to acoustic waves with wavelengths much shorter than the
radius of the Sun; therefore, we may ignore the curvature of the solar surface. This
allows us to make the simplifying assumption that the atmosphere is plane-parallel
with constant gravity pointing downward, $\bvec{g} = -g\bvec{\hat{z}}$. The height
coordinate $z$ is defined to increase upward.  We model the solar convection zone
with an isentropically stratified atmosphere that is excised above a height $z = -z_0$,
known as the truncation height.  Below this height, $z<-z_0$, the atmosphere is polytropic
and the gas pressure, mass density, and sound speed are power-law functions of height,

\begin{eqnarray}
	\rho_\e(z) &=& \rho_0 \left(-\frac{z}{z_0}\right)^m ,				\\
	P_\e(z) &=& \frac{g z_0 \rho_0}{m+1} \left(-\frac{z}{z_0}\right)^{m+1}
		 = P_0 \left(-\frac{z}{z_0}\right)^{m+1} ,				\\
	c_\e^2(z) &=& -\frac{gz}{m} .
\end{eqnarray}

\noindent In the preceding equations, the subscript ``e" is used to denote the
nonmagnetic atmosphere that is ``external" to the magnetic fibril. The constants
$\rho_0$ and $P_0$ are the surface values of the mass density and gas pressure. The
constant $m$ is the polytropic index, which has a special value that depends
on the ratio of specific heats $\gamma$, because the atmosphere is isentropically
stratified, $m = 1/(\gamma-1)$.

An extremely diffuse and hot gas exists above the truncation height. For this work,
we adopt the limiting case of a vacuum ($\rho \to 0$) with infinite temperature
($T \to \infty$), such that the pressure remains finite. At the surface, where
$z=-z_0$, the pressure of this ``hot corona" matches the pressure of the lower
atmosphere (the model convection zone).

Following BHCC96, we specify the truncation depth $z_0$ and the surface mass density
$\rho_0$ (and therefore the surface pressure $P_0$) by matching the surface layer
of our model, $z=-z_0$, with the $\tau_{5000} = 1$ level of a solar model by
\cite{Maltby:1986}.  At this layer in the solar model $g=2.775 \times 10^4$ cm s$^{-1}$,
$\rho_0 = 2.78 \times 10^{-7}$ g cm$^{-3}$, and $P_0 = 1.21 \times 10^5$ g cm$^{-1}$ s$^{-1}$.
We adopt $m=1.5$ (appropriate for $\gamma=5/3$), which is a reasonable approximation
to the bulk of the convection zone. With this choice of polytropic index, the truncation
depth and surface sound speed become $392$ km and 8.52 km s$^{-1}$, respectively.

\subsection{The Magnetic Fibril}
\label{subsec:equil_tube}

For simplicity, we assume that the magnetic fibril threading the truncated polytrope
obeys the following properties:

\begin{enumerate}
	\item The flux tube is straight, untwisted and axisymmetric.
	\item The tube has a circular cross section.
	\item The tube is vertically aligned and coincident with the
		 $\bvec{\hat{z}}$-axis.
	\item The tube has a potential magnetic field lacking internal currents.
	\item The boundary of the flux tube is marked by a current sheet.
	\item The tube is thin, meaning that its radius is significantly smaller than
		any scale length in the atmosphere, including the density scale height
		and the wavelength of incident acoustic waves.
\end{enumerate}

The last of these assumptions bears some further discussion. A sufficiently thin
flux tube is unable to support internal forces and structures beyond hydrostatic
balance. Thus, the gas pressure imposed by the external medium at the flux tube boundary
must be matched by a total pressure that is constant with radius across the tube.
This same argument holds for temperature as well. The temperature inside and outside
must be identical (as are the pressure scale heights), because any initial difference
would be destroyed by thermal and radiative diffusion. Continuity of total pressure
across the flux tube interface requires that the magnetic pressure also has
the same scale height.  Hence, the value of the plasma parameter $\beta = 8\pi P/B^2$
must be constant with height within the tube.

This simple statement of constant $\beta$ within the tube is a strong constraint on
the geometry of the flux tube.  Since the gas pressure in the solar atmosphere decreases
rapidly with height, the magnetic field strength within the tube also decreases rapidly,
and the flux tube must flare dramatically near the surface. Thus, at some height in
the atmosphere, the approximation that the tube is thin must break down. The magnetic
field outside of sunspots is generally found in the form of small magnetic elements.
For such elements, the flaring that occurs with height becomes important within the
chromosphere where neighboring flux elements begin to collide, thereby forming a magnetic
canopy. The thin-flux-tube approximation cannot be applied consistently within the upper
atmosphere.  In this work, we circumvent this problem by truncating the atmosphere and
only studying wave propagation in the region where the flux tube remains thin. As will
be discussed more fully in \S 4.3, the details of the tube geometry and thermodynamic
structure within the upper atmosphere can be mimicked by an appropriate choice of boundary
condition at the surface, $z = -z_0$.

For thin flux tubes of the sort described above, radial variation of the field strength
can be ignored and the flux tube is completely defined by its total magnetic
flux $\Theta$ and the plasma $\beta$. The tube's internal gas pressure $P(z)$, mass
density $\rho(z)$, axial field strength $B(z)$, and cross-sectional area $A(z)$
are given by the equations

\begin{eqnarray}
	P(z) &=& \frac{\beta}{\beta+1} P_\e(z) ,					\\
	\rho(z) &=& \frac{\beta}{\beta+1} \rho_\e(z) ,					\\
        \frac{B^2(z)}{8\pi} &=& \frac{1}{\beta+1} P_\e(z) ,				\\
	A(z) &=& {\displaystyle \frac{\Theta}{B(z)}} =
		 \left(\frac{\beta+1}{8\pi P_\e(z)}\right)^{1/2} \Theta .  \label{eqn:area}
\end{eqnarray}

\noindent For a flux tube with $\beta = 1.0$ embedded in a $m=1.5$ polytrope with a surface
mass density of $\rho_0 = 2.78 \times 10^{-7}$ g cm$^{-3}$, the magnetic field strength at
the surface, $z = -z_0$, is 1.2 kG.


\section{The $p$-Mode Oscillations}
\label{sec:pmodes}
\setcounter{equation}{0}

The nonmagnetic atmosphere surrounding the magnetic fibril supports acoustic oscillations
that are trapped in a wave guide just below the surface. The upper boundary reflects
upward-propagating waves and the increasing sound speed with depth refracts
downward-propagating waves back towards the surface. Since the atmosphere is isentropic, the
acoustic modes of this waveguide---the $f$ and $p$ modes---have a particularly simple
form and can be expressed using a displacement potential $\Phi$,

\begin{eqnarray}
	\bvec{\xi}_\e &=& \bvec{\nabla} \Phi ,    		    		\label{eqn:xi}	\\
	\delta P_\e &=& -\rho_\e \dparn[\Phi]{t}{2} ,			\label{eqn:dP}	\\
	\delta \rho_\e &=& -\frac{\rho_\e}{c_\e^2} \dparn[\Phi]{t}{2}. \label{eqn:drho}
\end{eqnarray}

The vector $\bvec{\xi}_\e$ is the fluid displacement, $\delta P_\e$ is the perturbed
gas pressure, $\delta \rho_\e$ is the perturbed mass density and $c_\e$ is the sound
speed. Since the tube is thin, it is incapable of supporting a temperature differential
with its surrounding; thus, the sound speed inside and outside the tube are the same
function of height $c(z) = c_\e(z)$. We drop the subscript on $c_\e$ from here on.

\subsection{Wave Equation}
\label{subsec:pmode_waves}

Equations \eqnref{eqn:xi}--\eqnref{eqn:drho} can be combined into a single partial differential
equation for the displacement potential,

\begin{equation}
	\dparn[\Phi]{t}{2} = c^2 \nabla^2\Phi - g\dpar[\Phi]{z} .	\label{eqn:wave_eqn}
\end{equation}

\noindent Equation \eqnref{eqn:wave_eqn} supports plane-wave solutions of the form

\begin{equation}
	\Phi({\bf x},t) = {\cal A} e^{-i\omega t} e^{ikx} Q(z) ,	\label{eqn:time_dependence}
\end{equation}

\noindent where ${\cal A}$ is the complex wave amplitude, $\omega$ is the temporal frequency,
$k$ is the wavenumber, and $Q(z)$ is the vertical eigenfunction; $Q(z)$ is dimensionless
and ${\cal A}$ has units of length squared. We have assumed without loss of generality
that the acoustic wave propagates in only one horizontal direction, the $x$-direction.
Direct substitution into equation \eqnref{eqn:wave_eqn} produces an ODE for the
eigenfunction $Q(z)$,

\begin{equation}
	\left\{c^2 \dern{z}{2} - g \der{z} + \left(\omega^2-k^2c^2\right)\right\}Q(z) = 0 .
\end{equation}

This equation can be cast in a useful dimensionless form by the substitutions

\begin{displaymath}
	\nu^2 \equiv \frac{m\omega^2 z_0}{g}, \qquad  w \equiv -2kz , \qquad \lambda \equiv 2kz_0 , \qquad \kappa \equiv {\displaystyle \frac{\nu^2}{\lambda}} , 
\end{displaymath}

\noindent resulting in the equation

\begin{equation}
	\left\{\dern{w}{2}+\frac{m}{w} \der{w}+\left(\frac{\kappa}{w}-
		\frac{1}{4}\right)\right\} Q(w) = 0 .			\label{eqn:ODE}
\end{equation}

The two solutions to equation~\eqnref{eqn:ODE} involve Whittaker's functions $W_{\kappa,\mu}(w)$
and $M_{\kappa,\mu}(w)$ \citep[see][]{Abramowitz:1964},

\begin{equation}
	Q(w) = w^{-\left(\mu+1/2\right)} \left\{ {W_{\kappa,\mu}(w)}\atop{M_{\kappa,\mu}(w)} \right. ,
\end{equation}

\noindent where $\mu = (m+1)/2$. Whittaker's $W$ function vanishes as $w\to\infty$,
but is poorly behaved at the origin $w=0$, whereas Whittaker's $M$ function is well
behaved at the origin and diverges exponentially as $w\to\infty$.

\subsection{$p$-Mode Boundary Conditions}
\label{subsec:pmode_BC}

As boundary conditions we require that the solution vanishes deep in the atmosphere
as $w\to\infty$, and we require that the Lagrangian pressure perturbation vanishes at
the free upper surface $z=-z_0$ (or equivalently, $w=\lambda=2kz_0$). The first of
these requirements makes Whittaker's $M$ function unsuitable. The second of these
boundary conditions places a restriction on the allowed values of $\kappa$ (or
equivalently $\lambda$).  Therefore, for a fixed value of the frequency, the
eigenvalue $\kappa=\kappa_n$ is quantitized and satisfies the following set of
equations, which simply restate that the Lagrangian pressure perturbation vanishes
at the surface,

\begin{equation}
	\left(\der{w} + \frac{\kappa_n}{m}\right)Q_n(\lambda_n) = 0 ,
\end{equation}

\noindent where,

\begin{equation}
	Q_n(w) = w^{-\left(\mu+1/2\right)} W_{\kappa_n, \mu}(w) .			\\
\end{equation}

\noindent After using recursion relations for Whittaker's $W$ function
\citep[see][]{Abramowitz:1964}, the last two equations can be reduced to

\begin{equation}
	W_{\kappa_{n},\mu+1}(\lambda_n) = 0 .	 \label{eqn:pmode_BC}
\end{equation}

\noindent The lowest order solution with $n=0$ is the fundamental mode or $f$ mode,
whereas the higher overtones $n>0$ correspond to the $p$ modes.

\subsection{Acoustic Mode Energy}
\label{subsec:pmode_energy}

The energy density of an acoustic wave propagating through an isentropic media is
given by \cite{Bray:1974},

\begin{equation}
	{\cal E} = \frac{1}{2}\rho_\e \left|\bvec{\dot{\xi}}_\e\right|^2 +
			\frac{1}{2}\frac{\delta P_\e^2}{\rho_\e c^2} ,	\label{eqn:Edensity}
\end{equation}

\noindent where the vector $\bvec{\dot{\xi}}_\e$ is the partial derivative of $\bvec{\xi}_\e$
with respect to time.

After taking temporal averages of equation \eqnref{eqn:Edensity}, integrating over
height, and multiplying by the surface area of the Sun, one finds the following
expression for the energy contained in $f$- and $p$-mode oscillations,

\begin{eqnarray}
	E_n &=& 4\pi R_\sun^2 \frac{g\rho_0}{4m} \frac{\nu^4}{\lambda_n^m} 
		\frac{\left|{\cal A}_n\right|^2}{z_0^2} {\cal N}_n ,	\label{eqn:modeenergy}	\\
									\nonumber \\
	{\cal N}_n &\equiv& \int^\infty_{\lambda_n} dw \; \frac{w^m}{\kappa_n}
		\left\{ \left(\der[Q_n(w)]{w}\right)^2 +
		\left(\frac{\kappa_n}{w}+\frac{1}{4}\right)Q_n^2(w)\right\} .
\end{eqnarray}


\section{Excitation of Tube Waves}
\label{sec:tubewaves}
\setcounter{equation}{0}

The magnetic fibril is buffeted and driven by $f$-mode and $p$-mode waves within
the convection zone. Magnetosonic oscillations of thin flux tubes have been examined
in detail by many authors \citep{Spruit:1981, Spruit:1984, Stix:1991, Ryutova:1993,
Bogdan:1996}, and can be fully described by two types of waves: kink and sausage.
Sausage waves are axisymmetric pressure pulses that produce displacements that 
are primarily parallel to the magnetic field, $\xi_\parallel(z,t)$. Kink waves
produce perpendicular displacements, $\xi_\perp(z,t)$, and magnetic tension and
buoyancy are the primary restoring forces.

\subsection{Thin-Flux-Tube Equations}
\label{subsec:tubewave_equations}
The sausage waves and the kink waves are driven respectively by the overpressure
and the transverse velocity imposed on the outer surface of the flux tube by
incident acoustic waves. Using the formulation of BHCC96, the sausage and kink waves
can be described by the equations

\begin{eqnarray}
	\left\{ \dparn{t}{2} - c_{\rm T}^2\dparn{z}{2} +
		\frac{\gamma g}{2}\frac{c_{\rm T}^2}{c^2}\dpar{z} \right\} \xi_\parallel &=&
		\frac{\rho_\e}{\rho} \frac{c_{\rm T}^2}{V_{\rm A}^2}
			\frac{\partial^3\Phi}{\partial z\partial t^2},
				\label{eqn:sausage}			\\
								\nonumber \\
	\left\{ \dparn{t}{2} - c_{\rm K}^2\dparn{z}{2} +
		\frac{\gamma g}{2}\frac{c_{\rm K}^2}{c^2}\dpar{z} \right\} \xi_\perp &=&
		2 \frac{\rho_\e}{\rho} \frac{c_{\rm K}^2}{V_{\rm A}^2}
			\frac{\partial^3\Phi}{\partial x\partial t^2},
				\label{eqn:kink}
\end{eqnarray}

\noindent where $V_{\rm A}$ is the Alfv\'en speed, $c_{\rm T}$ is the cusp or tube
speed, and $c_{\rm K}$ is the kink speed. This later speed is also referred
to as the ``mean" Alfv\'en speed, since it is the density-weighted mean of the Alfv\'en
speeds inside and outside the tube:

\begin{displaymath}
	V_{\rm A}^2 = \frac{B^2}{4\pi \rho} , \qquad
	c_{\rm T}^2 = \frac{c^2 V_{\rm A}^2}{c^2 + V_{\rm A}^2} ,  \qquad
	c_{\rm K}^2 = \frac{B^2}{4\pi (\rho + \rho_\e)}.
\end{displaymath}

Equations \eqnref{eqn:sausage} and \eqnref{eqn:kink}
describe the forced oscillations of sausage and kink waves on a slender
tube.  Clearly, in the limit of low $\beta$, the sausage waves are simply slow magnetosonic
waves propagating along the thin tube (see \citeauthor{Roberts:1978} [\citeyear{Roberts:1978}] for a derivation in an
atmosphere with generic stratification).  In this same limit, the kink waves are fast
magnetosonic waves, with an enhanced effective density due to the fact that transverse
motions of the tube must push and pull external fluid. However, in the parameter regime that
corresponds to photospheric magnetic elements, the plasma $\beta$ is roughly unity and
the distinction of fast versus slow magnetosonic waves proves less useful. In any event,
the sausage wave is primarily a pressure wave, whereas the kink wave is principally a
tension wave.

The shaking of the fibril by $p$-mode oscillations appears as the forcing on the right-hand
sides of equations \eqnref{eqn:sausage} and \eqnref{eqn:kink} where the pressure
perturbation and horizontal displacement have been replaced by appropriate derivatives
of the displacement potential. The forcing for the sausage waves arises from the requirement
that the total pressure is continuous across the tube boundary.  The kink waves are driven
by horizontal motions of the external media.

These equations have similar form and can be written compactly using the notation,
 
\begin{equation}
	\left\{ \dern{s}{2} + \frac{\mu+1}{s}\der{s} + \frac{\nu^2 \epsilon_\sigma}{s} \right\} \xi_\sigma
		= \frac{ {\cal A}_n}{z_0} f_\sigma(s) ,			\label{eqn:tubewaves}
\end{equation}

\noindent where $s = -z/z_0$ is a dimensionless depth and $\sigma$ represents either
$\parallel$ or $\perp$, corresponding to the sausage and kink waves,
respectively. The quantity $\epsilon_\sigma$ takes on the two values,

\begin{displaymath}
	\epsilon_\parallel = \frac{2m + \beta(m+1)}{2m}, \qquad	
		\epsilon_\perp = \frac{(m+1)(2\beta+1)}{2m} .
\end{displaymath}

\noindent The forcing function $f_\sigma(s)$ depends on whether we are driving
kink or sausage waves,

\begin{eqnarray}
	f_\parallel(s) &\equiv& -\frac{(m+1)(\beta+1)}{2m} \frac{\nu^2}{s}\der[Q_n(s)]{s} , \label{eqn:driver0}	\\
											    \nonumber		\\
	f_\perp(s) &\equiv& i \frac{(m+1)(\beta+1)}{2m} \frac{\lambda_n \nu^2}{s} Q_n(s) .  \label{eqn:driver1}
\end{eqnarray}

The homogeneous solutions to equation \eqnref{eqn:tubewaves} can be expressed in terms
of Hankel functions,

\begin{eqnarray}
	\psi_\sigma(s) &=& s^{-\mu/2} H_\mu^{(1)}\left(2\nu\sqrt{\epsilon_\sigma s} \right),	\\
										\nonumber	\\
	\theta_\sigma(s) &=& s^{-\mu/2} H_\mu^{(2)}\left(2\nu\sqrt{\epsilon_\sigma s} \right) = \psi_\sigma^*(s) .
\end{eqnarray}

\noindent For the chosen time dependence (see eq.~[\ref{eqn:time_dependence}]), the
$\psi_\sigma$ solution represents a downward-propagating wave, while
the $\theta_\sigma$ solution is an upward-propagating wave. Note that the two solutions
are simply complex conjugates of each other, thus allowing all further equations to
be expressed using solely $\psi_\sigma$. 

The solution to the driven problem is constructed from the homogeneous solutions
using a Green's function formulation,
	

\begin{equation}
	\xi_\sigma(s) = -\frac{i\pi}{2} \frac{ {\cal A}_n}{z_0}
		\Bigg\{ \psi_\sigma(s) \left[\Omega_\sigma + \int_1^s dr \; r^{\mu+1}
			\psi_\sigma^*(r)f_\sigma(r)\right]
			+ \theta_\sigma(s) \int_s^\infty dr \; r^{\mu+1}
			\psi_\sigma(r) f_\sigma(r) \Bigg\} .
				\label{eqn:tubewave_sol}
\end{equation}

This solution was constructed to satisfy specific boundary conditions. Deep in the
atmosphere, as $s\to\infty$ (or $z\to -\infty$), $\xi_\sigma$ becomes proportional to
$\psi_\sigma$, thereby satisfying a causal radiation condition. At the surface $s=1$
(or $z=-z_0$), the solution is a mixture of upward- and downward-propagating waves. The
exact proportion of $\psi_\sigma$ to $\theta_\sigma$ in this mixture is completely
specified by the boundary-condition parameter $\Omega_\sigma$.  The boundary condition
represents both the reflection that occurs directly at the upper surface and
the combined effect of the entire upper atmosphere $z > -z_0$, which is not included
explicitly in the calculation.  Any physical choice of boundary condition can be satisfied
by an appropriate value of $\Omega_\sigma$. For example, $\Omega_\sigma = 0$ enforces a
radiation condition, where only the upward-propagating wave is present at the surface.

The integrals that appear in equation \eqnref{eqn:tubewave_sol} form the basis for all
of the derived quantities that appear later. The $p$-mode eigenfunctions $Q_n$ are purely
real functions, and therefore the driving functions $f_\sigma$ are either purely real
(for the sausage) or purely imaginary (for the kink). This allows us to express both
integrals compactly for both the sausage mode and the kink mode,

\begin{eqnarray}
	{\cal J}_\sigma(s) &\equiv& \int_1^s dr \; r^{\mu+1} \psi_\sigma(r) f_\sigma(r),	\\
	{\cal I}_\sigma \equiv \lim_{s\to\infty} {\cal J}_\sigma(s) &=& \int_1^\infty dr \; 
			r^{\mu+1} \psi_\sigma(r) f_\sigma(r),					\\
	{\cal I}_\sigma - {\cal J}_\sigma(s) &=& \int_s^\infty dr \; r^{\mu+1} \psi_\sigma(r) f_\sigma(r).
\end{eqnarray}

We call ${\cal I}_\sigma$ the interaction integral between the $p$ mode and the respective
tube wave. Using these definitions explicitly, the displacements for the two types of waves
take on the forms

\begin{eqnarray}
	\xi_\parallel(s) &=& -\frac{i\pi}{2} \frac{{\cal A}_n}{z_0} \Bigg\{
		\psi_\parallel(s)   \Big[ \Omega_\parallel + {\cal J}_\parallel^*(s) \Big]
	   	+ \psi_\parallel^*(s) \Big[ {\cal I}_\parallel - {\cal J}_\parallel(s) \Big]
		\Bigg\},					\label{eqn:xipara}			\\
	\xi_\perp(s) &=& -\frac{i\pi}{2} \frac{ {\cal A}_n }{z_0} \Bigg\{
		\psi_\perp(s)   \Big[ \Omega_\perp - {\cal J}_\perp^*(s) \Big]
	   	+ \psi_\perp^*(s) \Big[ {\cal I}_\perp - {\cal J}_\perp(s) \Big]
		\Bigg\} .					\label{eqn:xiperp}
\end{eqnarray}

\noindent Note the difference in sign in front of ${\cal J}_\sigma^*$ in the first
term in the curly braces. This difference arises because $f_\parallel$ is purely
real and $f_\perp$ is purely imaginary.

In equations~\eqnref{eqn:xipara} and \eqnref{eqn:xiperp} the term $\psi_\sigma {\cal J}_\sigma^*$
represents the upward-propagating wave generated by the driver, while the term
$\psi_\sigma^* \left[{\cal I}_\sigma - {\cal J}_\sigma \right]$ represents the downward wave
directly generated by the driver. The term $\psi_\sigma \Omega_\sigma$ is a downward-propagating
wave arising from reflection off the upper surface, plus any waves propagating downward from the upper
atmosphere through the upper surface (perhaps caused by a reflection in the chromosphere or corona).

Note that the energy flux of waves generated by the driver and propagating away from the driving
region is the same both upward and downward. This can be seen by examining only those terms in
equations~\eqnref{eqn:xipara} and \eqnref{eqn:xiperp} that represent wave components directly
generated by the driver. The downward component is evaluated as $s\to\infty$, and the upward
component at $s=1$,

\begin{eqnarray}
	\xi_\sigma^{\rm (down)} = \mp \frac{i\pi}{2} \frac{ {\cal A}_n }{z_0} \psi_\sigma {\cal I}_\sigma^*, 	\\
	\xi_\sigma^{\rm (up)} = - \frac{i\pi}{2} \frac{ {\cal A}_n }{z_0} \psi_\sigma^* {\cal I}_\sigma .
\end{eqnarray}

\noindent This is a useful fact that will be exploited later.

\subsection{Energy Flux of Tube Waves}
\label{subsec:tubewave_fluxes}

At any point along the tube, the energy flux is given by the following expression
\citep{Bray:1974},

\begin{equation}
	\bvec{F} = \left(\delta P + \frac{\delta\bvec{B} \cdot \bvec{B} }{4\pi}\right) \bvec{\dot{\xi}}
		- \frac{\delta\bvec{B}\cdot\bvec{\dot{\xi}}}{4\pi}\bvec{B} . 		\label{eqn:Eflux}
\end{equation}

\noindent The perturbed magnetic field for sausage waves is given by equation (A8)
in Appendix A of BHCC96. The perturbed field for the kink waves can be derived directly
from the induction equation of MHD,

\begin{eqnarray}
	\frac{\delta B_\parallel B}{4\pi} &=& \frac{2}{2+\gamma\beta}\delta P_\e +
		\frac{\gamma\beta}{2+\gamma\beta}\frac{B^2}{4\pi}\dpar[\xi_\parallel]{z} -
		g\left(\rho - \frac{\gamma\beta}{2+\gamma\beta}\rho_\e\right) \xi_\parallel ,
				\label{eqn:dBpara}			\\
	\delta B_\perp &=& B\dpar[\xi_\perp]{z} . \label{eqn:dBperp}
\end{eqnarray}

Direct substitution of equations \eqnref{eqn:xipara}, \eqnref{eqn:xiperp}, \eqnref{eqn:dBpara}
and \eqnref{eqn:dBperp} into equation \eqnref{eqn:Eflux} produces the
following vertical energy fluxes at the two boundaries. As $s\to\infty$,

\begin{eqnarray}
	F_\parallel &=& -\frac{\gamma\beta}{2+\gamma\beta} \frac{\pi g\rho_0\omega}{4(m+1)(\beta+1)}
		\frac{\left|{\cal A}_n\right|^2}{z_0^2}
		 \Big|\Omega_\parallel + {\cal I}_\parallel^*\Big|^2 s^{\mu+1},		  \\
										\nonumber \\
        F_\perp &=& \quad \quad \quad\, -\frac{\pi g\rho_0\omega}{4(m+1)(\beta+1)}
                \frac{\left|{\cal A}_n\right|^2}{z_0^2}
                 \Big|\Omega_\perp - {\cal I}_\perp^*\Big|^2 s^{\mu+1},
\end{eqnarray}

\noindent and at $s=1$,

\begin{eqnarray}
	F_\parallel &=& \frac{\gamma\beta}{2+\gamma\beta} \frac{\pi g\rho_0\omega}{4(m+1)(\beta+1)}
		\frac{\left|{\cal A}_n\right|^2}{z_0^2}
		 \left( \left|{\cal I}_\parallel\right|^2 - \left|\Omega_\parallel\right|^2 +{\cal S} \right),
										\label{eqn:upflux_0} \\
										\nonumber 	     \\
	F_\perp &=& \quad\quad\quad\, \frac{\pi g\rho_0\omega}{4(m+1)(\beta+1)}
		\frac{\left|{\cal A}_n\right|^2}{z_0^2}
		\left( \left|{\cal I}_\perp\right|^2 - \left|\Omega_\perp\right|^2 \right) ,
										\label{eqn:upflux_1}
\end{eqnarray}

\noindent where,

\begin{eqnarray}
	{\cal S} &=& -\frac{(m+1)(\beta+1)}{2m} \nu^2 {\cal Q}_n \Bigg\{
		H_\mu^{(1)}\left(2\nu\sqrt{\epsilon_\parallel}\right)
			\Big[\Omega_\parallel + {\cal I}_\parallel^*\Big]
	      	+ H_\mu^{(2)}\left(2\nu\sqrt{\epsilon_\parallel}\right)
			\Big[\Omega_\parallel^* + {\cal I}_\parallel\Big]
		\Bigg\},								\\
									\nonumber	\\
	{\cal Q}_n &=& Q_n(s=1) = Q_n(w=\lambda_n) .
\end{eqnarray}

Note that the flux passing both down and up the tube are functions of the boundary
condition that is applied at the surface $s=1$ through the $\Omega_\sigma$ parameter.
Furthermore, the sausage mode also possesses a term ${\cal S}$ in the energy flux at the
upper surface $s=1$ which arises because the driver, equation~\eqnref{eqn:driver0},
does not vanish at the surface. This term is essentially the $p$-mode eigenfunction
multiplied by the real part of the vertical displacement within the tube. Depending
on the phasing between these two quantities, this surface driving term can take on
positive or negative values.

Furthermore, the flux deep in the atmosphere $s\to\infty$ increases with depth
$F\sim s^{\mu+1}$, but it does so at the same rate that the cross-sectional area of
the tube decreases,

\begin{eqnarray}
	A(s) &=& \left( \frac{(m+1)(\beta+1)}{8\pi g z_0 \rho_0} \right)^{1/2} \Theta \; s^{-(\mu+1)} , \nonumber \\
 	     &=& A_{\rm s} \; s^{-(\mu+1)} .
\end{eqnarray}

\noindent Therefore, below those layers in which $p$ modes drive waves, the rate
at which energy passes down the tube is constant with depth,

\begin{eqnarray}
	\dot{E}_\parallel &=& -\frac{\gamma\beta}{2+\gamma\beta}
				\frac{\pi g\rho_0 \omega A_{\rm s}}{4(m+1)(\beta+1)}
				\frac{\left|{\cal A}_n\right|^2}{z_0^2}
				\Big|\Omega_\parallel + {\cal I}_\parallel^*\Big|^2,
									\label{eqn:downE0} \\
									\nonumber	   \\	
	\dot{E}_\perp     &=& \quad\quad\quad\,
		   	      - \frac{\pi g\rho_0 \omega A_{\rm s}}{4(m+1)(\beta+1)}
				\frac{\left|{\cal A}_n\right|^2}{z_0^2}
				\Big|\Omega_\perp - {\cal I}_\perp^*\Big|^2.
					\label{eqn:downE1}
\end{eqnarray}
	
\noindent In these two equations, $\dot{E}$ is negative for energy escaping the $p$-mode
cavity, or in other words, if the energy flux is downward.

At the upper surface $s=1$ a similar result is obtained,
				
\begin{eqnarray}
	\dot{E}_\parallel &=& -\frac{\gamma\beta}{2+\gamma\beta}
				\frac{\pi g\rho_0 \omega A_{\rm s}}{4(m+1)(\beta+1)}
				\frac{\left|{\cal A}_n\right|^2}{z_0^2}
				\left(\Big|{\cal I}_\parallel\Big|^2 - \Big|\Omega_\parallel\Big|^2
					+ {\cal S}\right),
									\label{eqn:upE0} \\
									\nonumber	 \\
	\dot{E}_\perp     &=& \quad\quad\quad\,
		   	      - \frac{\pi g\rho_0 \omega A_{\rm s}}{4(m+1)(\beta+1)}
				\frac{\left|{\cal A}_n\right|^2}{z_0^2}
				\left(\Big|{\cal I}_\perp\Big|^2 - \Big|\Omega_\perp\Big|^2 \right).
					\label{eqn:upE1}
\end{eqnarray}

\noindent In these expressions, $\dot{E}$ is once again negative for energy escaping
the $p$-mode cavity. However, in this case this requires that the flux is upward into
the upper atmosphere above our model convection zone.

\subsection{Boundary Conditions}
\label{subsec:tubewave_BC}

We apply two different boundary conditions at the surface and track the
results of each. The first boundary condition is the requirement that
the stress vanishes at the upper surface $s=1$. Of course, for this boundary
condition, the energy flux through the upper surface must be identically zero.
We include this boundary condition because it is the same boundary condition
adopted by BHCC96. The second boundary condition that we apply maximizes
the energy flux that passes up through the upper surface. This boundary condition
allows us to compute an upper limit on the amount of tube-wave energy that is driven
into the upper atmosphere.

\figone

\subsubsection{Stress-Free Condition}
\label{subsubsec:tubewave_BC_stressfree}

For the sausage mode, the stress-free condition is enforced by requiring that the
divergence of the displacement vector vanishes at the upper surface. This is
equivalent to setting the Lagrangian pressure perturbation to zero. BHCC96 have
shown that for the sausage waves,

\begin{equation}
	\bvec{\nabla \cdot \xi} = \frac{\beta}{2+\gamma\beta} \frac{1}{P}
		\left(-\delta P_\e +\frac{B^2}{4\pi} \der[\xi_\parallel]{z} +
			g\rho_\e \xi_\parallel \right) .
\end{equation}

After inserting $\delta P_\e$ and $\xi_\parallel$, and after substantial manipulation,
one finds that $\bvec{\nabla\cdot\xi}=0$ requires,

\begin{eqnarray}
	\Omega_\parallel &=& i\frac{(m+1)(\beta+1)}{m\pi} \nu^2 \frac{{\cal Q}_n}{{\cal H}_\parallel}
		-\frac{{\cal H}_\parallel^*}{{\cal H}_\parallel} {\cal I}_\parallel ,
									\label{eqn:SFCpara}	\\
									\nonumber		\\
	{\cal H}_\parallel &\equiv& \nu\sqrt{\epsilon_\parallel}
					H_{\mu+1}^{(1)}\left(2\nu\sqrt{\epsilon_\parallel}\right)
		+(\beta+1)(\mu+1) H_{\mu}^{(1)}\left(2\nu\sqrt{\epsilon_\parallel}\right) .
\end{eqnarray}

For the kink waves, BHCC96 imposed the boundary condition at $s=1$ that the magnetic tension
force vanishes or equivalently, that the second derivative of the displacement with respect
to height vanishes. We do not follow their example, for the simple reason that the horizontal
force equation for the kink mode possesses two other terms, the buoyancy force and the
external $p$-mode forcing. In equation~\eqnref{eqn:kink} the term with the second derivative
with respect to height arises from the magnetic tension, while the term on the left-hand side
with the single derivative with respect to height is the buoyancy force. The external forcing
is the right-hand side of the equation.  Neither the buoyancy nor external forcing vanish when
the tension force is zero.  However, setting the displacement itself to zero at the surface is
sufficient to specify a net horizontal force of zero.  Therefore, this is the proper stress-free
boundary condition that we apply.  Evaluation of equation \eqnref{eqn:xiperp} at $s=1$ reveals
that this stress-free boundary condition requires

\begin{equation}
	\Omega_\perp = -\frac{H_\mu^{(2)}\left(2\nu\sqrt{\epsilon_\perp}\right)}{H_\mu^{(1)}\left(2\nu\sqrt{\epsilon_\perp}\right)} {\cal I}_\perp .
				\label{eqn:SFCperp}
\end{equation}

Direct substitution of the values of $\Omega_\sigma$ given by equations~\eqnref{eqn:SFCpara}
and \eqnref{eqn:SFCperp} into the relevant energy-transfer rates, equations~\eqnref{eqn:upE0}
and \eqnref{eqn:upE1}, reveal that the energy flux through the upper surface is zero, as
expected.

\figtwo

\subsubsection{Maximal-Flux Condition}
\label{subsubsec:tubewave_BC_maxflux}

The maximal-flux boundary condition allows us to place an upper limit on the energy
contained within coronal-loop waves that are generated by $p$-mode forcing in the
solar convection zone. We specify the value of $\Omega_\sigma$ by maximizing the energy
flux shown in equations \eqnref{eqn:upflux_0} and \eqnref{eqn:upflux_1},

\begin{eqnarray}
	\Omega_\parallel &=& -\frac{(m+1)(\beta+1)}{2m} \nu^2 {\cal Q}_n
		H_\mu^{(2)}\left(2\nu\sqrt{\epsilon_\parallel}\right) ,		\\
							\nonumber		\\
	\Omega_\perp &=& 0 .
\end{eqnarray}

\noindent These values for $\Omega_\sigma$ are the only extrema in $F_\sigma$, and a simple
check confirms that the upward fluxes are maximized at these values. Furthermore, the flux
through the upper surface is identical to the downward flux for this boundary condition.

One might assume that the maximal-flux boundary condition is equivalent to applying
a radiation condition at $s=1$. However, this is only true for the kink wave. A radiating
upper boundary requires that $\Omega_\sigma=0$. If the $p$-mode eigenfunction vanished
at the upper surface, the maximal-flux condition would indeed be $\Omega_\parallel = 0$.
The fact that the driver is nonzero at the surface ensures that the radiation condition
and maximal-flux condition are different requirements.


\section{Results}
\label{sec:results}
\setcounter{equation}{0}

In the following two subsections we examine several properties obtained from the
energy fluxes derived in \S\ref{subsec:tubewave_fluxes}. In particular, in
\S\ref{subsec:results_fluxes} we present the energy flux of tube waves driven
into the upper atmosphere by the $p$-mode driving. In \S\ref{subsec:results_damping}
we examine the damping imposed on the $p$ modes themselves by the excitation of
tube waves.

An examination of the energy-loss rates, equations \eqnref{eqn:downE0}--\eqnref{eqn:upE1},
reveal that all can be computed directly
from the $p$-mode eigenfunctions $Q_n(w)$, the eigenvalues $\kappa_n(\nu)$, the
mode-energy integral ${\cal N}_n$, and the interaction integrals ${\cal I}_\parallel$
and ${\cal I}_\perp$. We compute all of these quantities using a shooting technique
with adaptive-step-size fifth-order Runge-Kutta integration. The eigenproblem is solved
iteratively at each dimensionless frequency $\nu$ and for all mode orders $n<8$. 
Subsequently, once the eigenfunction and eigenvalue have been established,
all integrals (${\cal N}_n$, ${\cal I}_\parallel$, and ${\cal I}_\perp$) are computed
during the final iteration of the numerical integrator.

\figthree

\figfour

\subsection{Atmospheric Fluxes of Tube Waves}
\label{subsec:results_fluxes}

In \S\ref{subsec:tubewave_fluxes} we derived, for a single flux tube, the energy
flux of tube waves that passes through the upper surface into the upper atmosphere.
Since we are interested in comparing the derived fluxes with energy fluxes observed
within the chromosphere or corona, the energy flux produced by a single flux
tube is not of particular relevance. Instead, we need the energy flux produced
collectively by all of the flux tubes that might contribute to the observations.
It is easiest and most direct if we examine the aggregate behavior of the entire
Sun. The total rate of energy deposition $\dot{E}_\sigma^{\rm tot}$ is simply the
deposition rate for a single tube $\dot{E_\sigma}$ multiplied by the number of tubes
on the solar surface $N$. If we assume that all of the flux tubes that pierce
the photosphere are identical (with the same value of $\beta$, magnetic flux $\Theta$,
etc.) then the number of flux tubes depends only on the surface cross-sectional 
area of a single tube $A_{\rm s}$ and the filling factor $f$. Using
$N = 4 \pi R_\sun^2 f/A_{\rm s}$ and equations \eqnref{eqn:upE0}--\eqnref{eqn:upE1},
one finds

\begin{eqnarray}
	\dot{E}_\parallel^{\rm tot} &=& -\frac{\gamma\beta}{2+\gamma\beta}
				\frac{\pi^2 g\rho_0 \omega f R_\sun^2}{(m+1)(\beta+1)}
				\frac{\left|{\cal A}_n\right|^2}{z_0^2}
				\left(\Big|{\cal I}_\parallel\Big|^2 - \Big|\Omega_\parallel\Big|^2
					+ {\cal S}\right),
									\label{eqn:Etot0}	\\
									\nonumber		\\
	\dot{E}_\perp^{\rm tot}     &=& \quad\quad\quad\, - \frac{\pi^2 g\rho_0 \omega f R_\sun^2}{(m+1)(\beta+1)}
				\frac{\left|{\cal A}_n\right|^2}{z_0^2}
				\left(\Big|{\cal I}_\perp\Big|^2 - \Big|\Omega_\perp\Big|^2 \right).
									\label{eqn:Etot1}
\end{eqnarray}

The final step required to convert these equations into a physically meaningful graph
is specification of the amplitudes ${\cal A}_n$ of the incident $p$ modes.  We use
measured $p$-mode energies to accomplish this.  Figure \ref{fig:modeenergy} shows $p$-mode
energies for harmonic degrees between 4 and 150 as measured by \cite{Komm:2003}. The
energies are largely a function of frequency alone and reach a peak value of roughly
$2.5 \times 10^{28}$ ergs at a frequency of 3.2 mHz. To either side of this peak the
energy falls off exponentially with frequency. In order to extrapolate to frequencies
outside the observed range, we fit these energies with a curve of the form

\begin{equation}
	E_{\rm fit}(\eta) = a\exp \left(
			-\frac{\sqrt{\left(\eta-\eta_0\right)^2 + \delta^2}}{d}\right) .
				\label{eqn:Efit}
\end{equation}

\noindent In this equation, $\eta$ is the cyclic $p$-mode frequency, $a$ is an amplitude, $\eta_0$
is the frequency of peak energy, $d$ is a wing decay rate, and $\delta$ is a core width.
Using a maximum-likelihood procedure we
find $a=3.71\times 10^{28}$ ergs, $\eta_0=3.16$ mHz, $\delta = 0.152$ mHz, and $d = 0.372$ mHz.
This fit is shown in Figure \ref{fig:modeenergy} with the red curve.
With this fit we can estimate the $p$-mode amplitudes ${\cal A}_n$ by equating the mode
energies in equation \eqnref{eqn:modeenergy} with the fit,

\begin{equation}
	\left|{\cal A}_n\right|^2 = \frac{m\lambda_n^m}{\nu^4}\frac{z_0^2}{\pi R_\sun^2}
		\frac{E_{\rm fit}}{g\rho_0 {\cal N}_n} .
\end{equation}

With this final ingredient we can evaluate the energy-deposition rates appearing in
equations \eqnref{eqn:Etot0} and \eqnref{eqn:Etot1}. These rates represent the amount
of energy that is driven into the upper atmosphere in the form of sausage and kink waves.
Figure~\ref{fig:Eatmo} presents the resulting energy-deposition rates for $\beta = 1.0$.
Figure~\ref{fig:Eatmotot} shows the rate summed over all $p$ modes for three different
values of $\beta$, representing weak ($\beta =10$), intermediate ($\beta = 1.0$), and
strong magnetic field ($\beta = 0.1$). A value of $\beta=1.0$ is appropriate for small
magnetic elements in the photosphere, corresponding to a surface magnetic field strength
of 1.2 kG. The other values of $\beta$ correspond to surface magnetic field strengths
of 0.53 kG for $\beta = 10$ and 1.6 kG for $\beta = 0.1$. We have included the results
for all three values of $\beta$ to demonstrate the behavior as $\beta$ changes.  In
particular, note that as the surface field strength decreases ($\beta$ increases) the
energy-transfer rate for kink waves decreases, whereas the rate for sausage waves increases.
For all values of $\beta$ the falloff at both high and low frequency is the result of
the decrease in the measured $p$-mode energy spectrum (and hence amplitude ${\cal A}_n$) away from
the peak at 3.2 mHz.

\subsection{$p$-Mode Damping Rates}
\label{subsec:results_damping}

Since the tube waves carry away energy from the acoustic cavity, this energy must
come from the $p$ modes themselves, thereby damping the solar acoustic modes. We
define a damping rate separately for the excitation of sausage and kink waves,

\begin{equation}
	\Gamma_\sigma = -\frac{1}{2\pi} \frac{\dot{E}_\sigma}{E_n} .
\end{equation}

Using the energy-deposition rates provided in equations \eqnref{eqn:downE0}--\eqnref{eqn:upE1},
we compute the damping rates,

\begin{eqnarray}
	\frac{\Gamma_\parallel}{\omega} &=& \frac{\beta}{4(\beta+1)\epsilon_\parallel}
		\frac{A_{\rm s}}{4\pi R_\sun^2} \frac{\lambda_n^m}{\nu^4}
		\frac{ \Big| \Omega_\parallel + {\cal I}_\parallel^* \Big|^2
		+ \Big|{\cal I}_\parallel\Big|^2 - \Big|\Omega_\parallel\Big|^2 + {\cal S} }{{\cal N}_n}, \\
												\nonumber	\\
	\frac{\Gamma_\perp}{\omega} &=& \frac{1}{2\gamma(\beta+1)}
		\frac{A_{\rm s}}{4\pi R_\sun^2} \frac{\lambda_n^m}{\nu^4}
		\frac{ \Big| \Omega_\perp - {\cal I}_\perp^* \Big|^2 
		+ \Big|{\cal I}_\perp\Big|^2 - \Big|\Omega_\perp\Big|^2 }{{\cal N}_n} .
\end{eqnarray}

\noindent In each of the expressions above, the first term in the numerator of the final fraction corresponds
to energy lost down the tube, while the remaining terms in the numerator arise from energy lost through
the upper surface. As stated previously, for the stress-free boundary condition the
downward term is identically zero, and for the maximal-flux boundary condition the upward and downward
terms are equal.

Figures~\ref{fig:gamma_beta0.1}--\ref{fig:gamma_beta10} present these damping rates
for three values of $\beta$. The most unusual feature within these graphs occurs for
$\beta = 0.1$ in Figure~\ref{fig:gamma_beta0.1}. At a frequency around 5.5 mHz, the
damping rate arising from the excitation of kink waves with the stress-free boundary
has nulls where the damping drops to zero. Such nulls are to be expected.  First, as
previously noted, the upward direct wave and the downward direct wave generated
by the driver possess the same energy flux. Second, the stress-free boundary reflects
all upward-propagating wave energy back into the convection-zone model. These two facts
lead one to the conclusion that the reflected wave and the downward direct wave have
the same amplitude.  There will be frequencies at which the phases of these two components
are such that complete destructive interference occurs. At such a frequency, the damping
rate will vanish because energy is driven neither up nor down the tube. A similar result
was found by \cite{Crouch:1999}.


\section{Discussion}
\label{sec:discussion}
\setcounter{equation}{0}

We have computed the energy flux  of sausage and kink waves that are generated in the
solar convection zone by the buffeting of magnetic fibrils by ambient $f$ and $p$ modes.
We treat the fibrils as vertically-aligned thin flux tubes. The tubes act as waveguides,
much like a fiber-optic cable that ducts the propagation of light. Waves are generated locally
within the $p$-mode cavity and freely propagate both up and down the tubes---one can easily
verify that there are no turning points for our model's stratification. Those waves that
propagate downward extract energy from the incident $p$ mode and disappear into the
convection zone. Those waves that propagate upward also damp the $p$ mode and travel
through the photosphere along the flux tube into the upper atmosphere. Such waves may
well be the source of both coronal-loop oscillations and the ubiquitous waves observed
throughout the corona.

\figfive

\subsection{Energy Flux of Tube Waves in the Upper Atmosphere}
\label{subsec:discussion_Eflux}

To determine if $p$-mode buffeting of small magnetic elements is a viable mechanism
to explain the waves observed in the corona, we directly compare our derived wave fluxes
with the observed fluxes. First, we comment that the spectral behavior of our derived
energy fluxes is correct by construction. As stated previously, the shape of observed
coronal wave spectra mimics spectra of $p$-mode energies---a broad peak at 3.5 mHz without
resonances or nulls. Since our energy fluxes are directly proportional to the $p$-mode
energy spectra, we clearly match this observed property well. Now, to determine whether the
predicted amount of wave energy is reasonable. Table 1 provides the results of summing the
tube-wave fluxes generated by all $p$ modes with a frequency below 5 mHz. In different
columns, we include both the energy flux (ergs cm$^{-2}$ s$^{-1}$) and energy-transfer
rate (ergs s$^{-1}$) for the entire solar surface.

\bigskip
\centerline{\small Table 1: INTEGRATED ATMOSPHERIC ENERGY FLUX}
\bigskip

\begin{center}
{\footnotesize
\begin{tabular}{cccccc}

{$\beta$} & {$\dot{E}_\parallel/f$} & {$\dot{E}_\parallel/(4\pi R_\sun^2 f)$} & {$\dot{E}_\perp/f$} & {$\dot{E}_\perp/(4\pi R_\sun^2 f)$}\\
{}        & {\{ergs s$^{-1}$\}}            & {\{ergs cm$^{-2}$ s$^{-1}$\}}                     & {\{ergs s$^{-1}$\}}        & {\{ergs cm$^{-2}$ s$^{-1}$\}} \\
\hline
0.10 & $3.81 \times 10^{27}$ & $6.25 \times 10^{4}$ & $1.01 \times 10^{29}$ & $1.65 \times 10^{6}$ \\
1.00 & $3.54 \times 10^{28}$ & $5.82 \times 10^{5}$ & $5.41 \times 10^{28}$ & $8.88 \times 10^{5}$ \\ 
10.0 & $1.84 \times 10^{29}$ & $3.02 \times 10^{6}$ & $1.35 \times 10^{28}$ & $2.22 \times 10^{5}$ \\
\hline
\end{tabular}

}
\end{center}

The energy fluxes presented in Table 1 are quite significant. \cite{DeMoortel:2007}
estimate that, depending on the mass density of coronal loops, the energy carried by loop
oscillations is $\dot{E} \approx f_{\rm c} \times 3 \times (10^{25}-10^{26})$ ergs s$^{-1}$,
where $f_{\rm c}$ is the coronal filling factor. \cite{Tomczyk:2007} estimate that the
upward energy flux of waves in the high corona is 10 ergs cm$^{-2}$ s$^{-1}$, or
$\dot{E} \approx 10^{24}$ ergs s$^{-1}$ if multiplied by the solar surface area.

If we adopt a typical value of $\beta=1.0$ for magnetic elements in the photosphere
and a value of $f = 0.01$ for the photospheric filling factor, we find that our derived
energy-transfer rates are $3.5 \times 10^{26}$ ergs s$^{-1}$ and $5.4 \times 10^{26}$ ergs s$^{-1}$
for the sausage and kink waves, respectively. While these numbers slightly exceed the
estimates of \cite{DeMoortel:2007} and exceed the estimate of \cite{Tomczyk:2007}
by several orders of magnitude, we must keep in mind that our theoretically derived
fluxes are the maximum possible flux that can be driven upward at the photospheric
level. Reflection and absorption within the chromosphere and transition region are
ignored. Clearly sufficient energy can be converted from $p$ modes to tube waves to
explain the observations; however, whether the attenuation of such waves allows them
to reach high into the corona remains to be seen. We remind the
reader that flux tubes can act as waveguides. The acoustic cutoff frequency of the
nonmagnetized atmosphere is not the relevant quantity to determine whether the tube
waves are propagating. Instead, the sausage and the kink waves have separate cutoff
frequencies which can be lower than the acoustic cutoff of the surrounding media,
depending on the properties of the tube \citep[see][]{Roberts:1978,Musielak:2001}.
Therefore, inclination of the field may not be required for magnetosonic waves to
``tunnel" through the photosphere and temperature-minimum region.

\figsix

\subsection{Damping and $p$-Mode Line Widths}
\label{subsec:discussion_damping}

The excitation of tube waves extracts energy from the incident $p$ mode that drives
the oscillations. Figures~\ref{fig:gamma_beta0.1}--\ref{fig:gamma_beta10} show the resulting
damping rate. In order to determine whether these damping rates are significant, in
Figure~\ref{fig:gamma_data} we plot the damping rates for $\beta = 1.0$ and a filling factor
of 0.01. We then overlay the $p$-mode full widths measured by \cite{Komm:2003}.  From
this figure, one can see that the excitation of tube waves is likely to be a significant
damping source only for the lowest order $p$ modes. The damping is also a strong
function of the boundary condition. The stress-free boundary produces significant
damping only for the $f$ mode, and most of the energy is converted into kink waves.
For the maximal-flux boundary, markedly more energy is converted into sausage waves
than kink waves, and the damping is important for $n < 5$.

\figseven

\subsection{Conclusions}
\label{subsec:discussion_conclusions}

We have demonstrated that $p$-mode buffeting of thin flux tubes can easily generate
the flux and spectral dependence of waves observed in the corona, in the form of either
longitudinal sausage waves or transverse kink waves. However, we only show that these
large fluxes are possible at the photospheric level. Therefore, the remaining problem
is one of transmission: Can the tube waves successfully propagate high into the corona
without being absorbed or reflected at lower levels in the atmosphere? We plan in future
efforts to include a realistic model of the temperature minimum and low chromosphere,
thereby accounting for the flaring of flux tubes with height and their eventual merger
into a magnetic canopy.  However, the semianalytic method used here is likely to prove
of limited usefulness once the model atmosphere and model flux tubes become sufficiently
complicated. Numerical simulations of the interaction between $p$ modes and flux tubes
will probably provide the ultimate answer. Such models have been previously built to
investigate the absorption of acoustic waves by sunspots; however, these models have
by explicit construction eliminated upward propagation of the resulting tube waves.
The inclusion of a chromosphere and realistic upper boundary conditions is needed.


\acknowledgments

We thank Rudi Komm for providing data of $p$-mode energies and we thank D. Haber
and G. Dickinson for help with the preparation of this manuscript.
We are also indebted to the anonymous reviewer who made useful suggestions. B.W.H. acknowledges
support from NASA and NSF through grants NAG5-12491, NAG5-13520, NNG05GM83G, NNX07AH82G
and ATM-0219581. R.J. acknowledges discussions with V. Nakariakov and funding from
Engineering and Physical Sciences Research Council, grant EP/C548795/1.

\end{document}